\documentstyle[aps]{revtex}

\begin{document}
\draft
\title{The effect of aging on network structure}
\author{Han Zhu$^1$, Xinran Wang$^1$ and Jian-Yang Zhu$^{2,3\thanks{%
Author to whom correspondence should be addressed. Address correspondence to
Department of Physics, Beijing Normal University, Beijing 100875, China.
Email address: zhujy@bnu.edu.cn}}$}
\address{$^1$Department of Physics, Nanjing University, Nanjing, 210093, China\\
$^2$CCAST (World Laboratory), Box 8730, Beijing 100080, China\\
$^3$Department of Physics, Beijing Normal University, Beijing 100875, China}
\maketitle

\begin{abstract}
In network evolution, the effect of aging is universal: in scientific
collaboration network, scientists have a finite time span of being active;
in movie actors network, once popular stars are retiring from stage; devices
on the Internet may become outmoded with techniques developing so rapidly.
Here we find in citation networks that this effect can be represented by an
exponential decay factor, $e^{-\beta \tau }$, where $\tau $ is the node age,
while other evolving networks (the Internet for instance) may have different
types of aging, for example, a power-law decay factor, which is also studied
and compared. It has been found that as soon as such a factor is introduced
to the Barabasi-Albert Scale-Free model, the network will be significantly
transformed. The network will be clustered even with infinitely large size,
and the clustering coefficient varies greatly with the intensity of the
aging effect, i.e. it increases linearly with $\beta $ for small values of $%
\beta $ and decays exponentially for large values of $\beta $. At the same
time, the aging effect may also result in a hierarchical structure and a
disassortative degree-degree correlation. Generally the aging effect will
increase the average distance between nodes, but the result depends on the
type of the decay factor. The network appears like a one-dimensional chain
when exponential decay is chosen, but with power-law decay, a transformation
process is observed, i.e., from a small-world network to a hypercubic
lattice, and to a one-dimensional chain finally. The disparities observed
for different choices of the decay factor, in clustering, average node
distance and probably other aspects not yet identified, are believed to bear
significant meaning on empirical data acquisition.
\end{abstract}


\section{Introduction}

Recently, the computer-aided data acquisition has led to an explosion of
interest in probing the complex network systems\cite{review1,review2,review3}%
. So far, many of the various networks in reality, such as the World Wide Web%
\cite{WWW}, social networks\cite{social}, and biological networks\cite
{Hierar2}, etc., are believed to share the following characteristics\cite
{review1,review2,review3,Hierar1,ass-mix}: (1) a small, relative to their
large size, average distance between nodes; (2) a power-law degree
distribution, often followed by a truncation; and a (3) highly clustered,
(4) hierarchical and (5) correlated structure (explanations see below).
Besides, there are many other features that are often not as readily
apparent and much less well understood consequently.

Aimed at a theoretical description of these findings, the Watts-Strogatz
(WS) Small-World model\cite{WS} is useful for systems that are largely
regular, and presents properties (1) and (3) listed above. Meanwhile the
Barabasi-Albert (BA) Scale-Free model\cite{BA} satisfactorily characterizes
most of the networks where geological distance is not so important. It
considers growth and preferential attachment irrespective of distance, with
properties (1) and (2) as outcome. We may notice that, in this model,
properties (3) and (4) are still missing\cite{Hierar1}, and, as is shown
below, a neutral degree-degree correlation is predicted for large degrees
(the degree of a node is defined as the number of the its links). This may
be due to the highly-simplified assumptions of the model. To develope these
two well-established models, theoretical effort has just begun.

In the generalization of the BA model, many novel and realistic aspects have
been investigated in the past few years\cite{review1,review3}. It is now
known that degree distribution in reality may deviate from a pure power law.
According to the extent of the deviation, the distribution patterns may be
categorized into three groups\cite{aging-pro1}: scale-free (power law),
broad-scale (power law with a sharp cut-off) and single-scale
(fast-decaying). In order to explain this deviation, the effect of aging and
a consequent loss of activity have been introduced\cite
{aging-pro1,aging-pro2,aging-pro3,growth-deactivation}. This is a common
mechanism in reality\footnote{%
This effect can be, for example, pictured for the network of actors (the
actors are linked if they both appear in the cast of one film). The more
famous an actor is, the more chances he will have to act in new movies. But,
however famous he may be, every star will become gradually inactive as time
passes. This is also supported by the citation rate data of the years
1987-1998\cite{growth-deactivation,citationweb1}, as shown in Fig. 1. Except
for the first three years prior to the publication year, the citaiton rate
gradually decreases with age.} and, combined with the BA model, it results
in a tunable truncation of the power law\cite{aging-pro1,aging-pro2}, i.e.,
the degree distribution can be turned from scale-free with no aging, to
broad-scale with slow aging and the single-scale with fast aging. This
finding justifies further investigation of its influence on the structure
and function of networks, which is the aim of the present work.

This article is organized as follows: In Sec. \ref{Sec. 2}, we provide
empirical evidence and quantify the aging effect in a model, which reduces
to the BA network when the aging effect vanishes. In Sec. \ref{Sec. 3}, the
effect of aging on network structure is described in four aspects:
clustering (Sec. \ref{Sec. 3. 1}), hierarchical structure (Sec. \ref{Sec. 3.
2}), degree-correlation (Sec. \ref{Sec. 3. 3}) and the average distance
between nodes (Sec. \ref{Sec. 3. 4}). As the intensity of the aging effect
grows, in most cases we witness a continuous transformation of the network
structure, while there are also a few abrupt changes. Finally, Sec. \ref
{Sec. 4} is the summary with some discussions.

\section{The model}

\label{Sec. 2}As the first step, we give the definition of the model with a
tunable effect of gradual aging: At each time step, a newly added node is
attached to $m$ existing nodes, with the probability proportional (1) to the
degree $k$ of the considered node, as in the BA model and (2) to a simple
function $f\left( \tau \right) $, where $\tau $ is the age of the considered
node. Thus the evolution of the network can be approximately characterized
by the following equation, 
\begin{equation}
\frac{\partial k\left( t_i,T\right) }{\partial T}=\frac{mk\left(
t_i,T\right) f\left( T-t_i\right) }{\sum_tk\left( t,T\right) f\left(
T-t\right) },
\end{equation}
where $k\left( t_i,T\right) $ denotes the expected degree at time $T$ of the
node born at $t_i$, and $\sum_tk\left( t,T\right) f\left( T-t\right) $ is
the normalization factor.

If $f\left( \tau \right) =1$, the probability that an existing node receives
new links becomes solely proportional to its degree, and this model reduces
to the BA model\cite{BA}, 
\begin{equation}
\frac{\partial k\left( t_i,T\right) }{\partial T}=\frac{mk\left(
t_i,T\right) _i}{\sum_tk\left( t,T\right) }=\frac{k\left( t_i,T\right) }{2T},
\label{BA-evolution}
\end{equation}
where we have used the fact that the normalization factor $\sum_tk\left(
t,T\right) =2mT$. On the other hand, when $f\left( \tau \right) $ decays
fast enough, we assume that the normalization factor, $\sum_tk\left(
t,T\right) f\left( T-t\right) $, reaches an asymptotic nondivergent value in
the limit of infinite network size. Thus, the expected degree of a given
node grows as, 
\begin{equation}
\frac{\partial k\left( t_i,T\right) }{\partial T}=\frac{k\left( t_i,T\right)
f\left( T-t_i\right) }M,
\end{equation}
where $M=\frac 1m\sum_tk\left( t,T\right) f\left( T-t\right) $. To test its
validity and obtain the factor $f\left( \tau \right) $, we apply this model
to the scientific citation web\cite
{growth-deactivation,citationweb1,citationweb2,citationweb3,citationweb4}, a
rather complex network formed by the citation patterns of scientific
publications, with the nodes standing for published articles and a directed
edge representing a reference to a previously published article.

Fig. 1 shows the citation rate data of the years 1987-1998\cite
{growth-deactivation,citationweb1} obtained from the ISI\ database. The
number of papers published in each year (Fig. 1a) is approximately stable
(as we have assumed), and, in order to further get rid of the fluctuation,
we use a set of relative values, $F_{1987\rightarrow 1998}$, $%
F_{1988\rightarrow 1998}$,..., $F_{1998\rightarrow 1998}$, i.e. of the
papers published in 1987, 1988, ..., 1998, the fraction that is cited in
1998. Then we reinterpret them as the following: each time step corresponds
to a year, and as soon as a node is introduced in the $Y$th step, its
initial degree is taken as $F_{1998\rightarrow 1998}$; within step $\left(
Y+1\right) $, the degree increases by $F_{1997\rightarrow 1998}$; ... within
step $\left( Y+11\right) $, it increases by $F_{1987\rightarrow 1998}$, to
finally $\sum_{i=1987}^{1998}F_{i\rightarrow 1998}$. Thus we can
approximately obtain both $dk\left( t\right) /dt$ and $k\left( t\right) $ as
a function of time. Then $f\left( t\right) $ can be calculated as $\left[
1/k\left( t\right) \right] dk\left( t\right) /dt$, and it is shown to be
proportional to, approximately, $e^{-0.28t}$ (Fig. 1b). In the following we
shall choose the factor to be

\begin{equation}
f\left( \tau \right) =e^{-\beta \tau },
\end{equation}
where $\beta $ is a tunable parameter. This particular measurement does not
exclude other functions, such as $\tau ^{-\nu }$\cite{aging-pro2,aging-pro3}%
, as possible choices. They are also studied, and found to yield similar
results in most respects (while several interesting differences are
highlighted below).

In the following we show how structural properties can be changed by the
aging effect introduced in the way quantified above.

\section{Transformation of network structure}

\label{Sec. 3}

In this section, we study the transformation of network structure by the
aging effect. First of all we pay attention to the mostly studied property
of complex networks, the vertex degree distribution. It is well known that
the aging effect may result in a transformation of the degree statistics,
from scale-free with no aging, to broad-scale with slow aging and to
single-scale with fast aging\cite{review1,review3,aging-pro1,aging-pro2}.
The detailed analytical and numerical study of evolving networks with
power-law aging that supports this idea can be found in Ref. \cite
{aging-pro2}. Here, with exponential aging, we have also found by numerical
simulations a very similar process.

It may be interesting to turn to the empirical results of the citation web.
In Ref. \cite{citationweb3}, the network formed by articles citing each
other in Physical Review D has been studied and the degree distribution
significantly deviates from a power law in the range of relatively small
degrees. In Ref. \cite{citationweb4}, the study has been extended to the
out-degree distributions of the networks formed by articles in a variety of
journals. The distributions have a maximum at intermediate out-degrees,
followed by an exponential tail for large out-degrees (single scale). These
pictures actually can be reproduced by tuning the decay factor $\beta $ from
small to large. The similar effect can be found in some figures of Ref. \cite
{aging-pro1,aging-pro2}.

In the following, we report the investigation of network structure
transformation by aging effect in four aspects: clustering, hierarchical
structure, degree correlation, and average node distance\footnote{%
On completion of this work, we have noticed that in Ref. \cite{deactivation3}%
, Vazquez {\it et al.} have investigated the deactivation model in similar
aspects. Actually, the deactivation model, just like the present model, can
be viewed as a specific kind of aging effect. However here we study gradual
aging and in most cases a gradual transformation process is demonstrated by
tuning the parameter. At the same time, there are some important different
results, e.g. those concerning the small-world effect.}. The definitions and
a brief review of relevant results can be found in the head of each
subsection.

\subsection{Clustering}

\label{Sec. 3. 1}

A useful tool to characterize the network structure is the clustering
coefficient $C$, which is defined as the average probability that a pair of
nearest neighbors of a given node is also connected. For example, if the
node $i$ has $k_i$ links, and among its $k_i$ nearest neighbors there are $%
\varepsilon _i$ edges, then the clustering coefficient of the node $i$ is
defined by 
\[
C_i=\frac{2\varepsilon _i}{k_i\left( k_i+1\right) }. 
\]
The clustering coefficient of the whole network is given by the average
value. A common property of, for example, social networks is that cliques
form, i.e. friends of yours are much more likely to be friends of each other
than people selected at random, thus resulting in a high clustering
coefficient. In the BA model, with the highly simplified assumptions, $C$
decreases with system size $N$ as $\left( \ln N\right) ^2/N$\cite
{deactivation2}. It is significantly lower than actual measurement, which is
to a high degree unaffected by the system size\cite{review1}.

Our study of the aging effect on network structure begins with the
investigation of clustering. Fig. 2 shows several typical curves of the
clustering coefficient $C\left( N\right) $ as a function of network size $N$%
. As aging gradually grows, we observe an interesting transformation
process, which can be roughly separated into several stages. (1) $\beta =0$:
The network reduces to the BA model and $C$ decreases to finally $0$ as the
system grows. (2) $0\prec \beta \prec 10^{-3}$: Now a slight aging effect is
introduced, and $C$ is significantly lowered given a relatively small system
size, but, when the size grows, this gap is contracting. (3) $10^{-3}\prec
\beta \prec 10^{-1}$: As aging is becoming more and more manifest, $C$ is
greatly enhanced. Given a relatively small system size, it may still be
lower than the value obtained in a non-aged network of the same size.
However, with a much lower rate of decreasing, it quickly exceeds that
non-aged value after reaching a crosspoint. As the system grows larger, the
curve is becoming increasingly flat and $C$ finally approaches a stable
value. (4) As $\beta $ continues to grow, this asymptotic value is quickly
rising, as shown by the curve at $\beta =0.5$. After a certain peak is
reached, it quickly falls back to finally zero.

In the following we analyze our observations. In networks, clustering is
determined by a competition of orderliness and randomness. In the BA model,
newly added nodes are more likely to be linked to earlier introduced nodes,
which generally have more links. This orderliness, however, is weakened by
the increasing randomness as the network size grows, thus resulting in a
vanishing $C$. When the aging effect is considered, the old nodes gradually
lose their activity in network function and growth. For a newly added node,
it is more probable to be linked with a temporally closer node, thus forming
a chain-like structure on the large scale. The region which a given node may
be linked with is of a finite effective size, however large the whole
network may actually be. This may explain why the $C\left( N\right) $ curve
quickly reaches a stable value when the aging effect is considered.
Different from the prediction of the BA model, this stable value is finite
even with a vanishingly small $\beta $. Now we explain how and why this
value varies with the intensity of the aging effect. In an aged network,
while the global randomness as a result of the large network size is absent,
the orderliness that older nodes receive more links is also lacked. The
clustering coefficient of a given node is now determined by the structure of
its neighboring region. When the aging effect is very weak, the size of this
region is relatively large and there is an even distribution of the
probability that the considered node is connected with a given member of
this region. The randomness caused by such an even distribution results in a
small value of $C$. As the aging effect grows, the size of this region
contracts and the probability distribution becomes more concentrated. Thus
the randomness is inhibited, and $C$ is significantly enhanced. However,
with very strong aging effect, $C$ diminishes as the probability
distribution becomes increasingly centralized. Finally, $C$ approaches zero
when each node can be linked only with the node introduced right before it.

The analysis above can be quantified by an approximate calculation of the
asymptotic clustering coefficient, $C\left( \beta \right) $, as a function
of $\beta $. The details can be found in the Appendix and the result, 
\begin{equation}
C\left( \beta \right) =\frac{6m^3}{2m\left( 2m-1\right) }e^{-\beta }\frac{%
\left( 1-e^{-\beta }\right) ^3}{\left( 1-e^{-2\beta }\right) ^2},
\label{Cbeta}
\end{equation}
in comparison with the simulation is shown in Fig. 3. From Eq. (\ref{Cbeta})
we can see that $C\left( \beta \right) $ increases linearly with $\beta $
for small values of $\beta $, and decays as $e^{-\beta }$ for large values
of $\beta $.

When we compare the two decay factors, $e^{-\beta \tau }$ and $\tau ^{-\nu }$%
, there is an interesting observation. In a network of $10,000$ nodes, we
measure the average clustering coefficient $C_1$ of the first half and $C_2$
of the second half (each containing $5,000$ nodes) respectively. When $%
e^{-\beta \tau }$ is chosen, two sets of approximately equal values are
obtained; however, when $\tau ^{-\nu }$ is chosen, $C_2$ equals $C_1$ only
with strong or weak aging, and in the middle ground $C_2$ is significantly
lower (Fig. 4). This interesting disparity justifies further investigation
and measurement in the study of real networks.

This finding also reminds us that, on empirical data acquisition, the effect
of aging plays a crucial role, since researchers often have a limited access
to the whole system, and will probably consider the most recently grown
part. Whether such a limited investigation can correctly represent the
overall system may strongly depend on the type of the aging effect.

\subsection{Hierarchical structure}

\label{Sec. 3. 2}

\label{Sec. 3.2}Now we go beyond the average clustering coefficient and
calculate $C\left( k\right) $ as a function of $k$. Here $C\left( k\right) $
denotes the expected clustering coefficient of a node with $k$ degrees%
\footnote{%
The method of our simulation is as the following: In each of the many
independent runs, after a network is generated, the degree $k$ and the
clustering coefficient $C$ of each node is measured. Then we calculate the
average $C$ of the nodes that have degree $k$. Finally the results are
further averaged over the independent runs.}. For complex networks, this
relationship is often of much significance because it is a useful tool to
inspect the intrinsic hierarchy of the topology. In the following we briefly
discuss the physical ground and then present our results with the aging
effect.

In reality, networks are often fundamentally modular\cite{Hierar2,Hierar1}:
nodes have a tendency to combine into subgroups in which they are highly
interconnected but have relatively few links to nodes outside. For example,
in society such groups may represent families, and in WWW they can denote
communities with shared interests. Numerous such groups then constitute the
whole system in a hierarchical manner. In some way the network might look
like a fractal graph\cite{Hierar3} (see Fig. 1 of Ref. \cite{Hierar1}). This
structure can be characterized quantitatively by a simple scaling law: $%
C\left( k\right) \sim k^{-\gamma }$\cite{Hierar1,Hierar3}. The coefficient $%
\gamma $ has been measured to be approximately $0.75$ on the Internet at the
autonomous system level\cite{Internet1,Internet2}.

When no aging effect is considered, the BA model does not show such a
property and we expect to observe $C\left( k\right) $ as a horizontal line
subject to fluctuations\cite{Hierar1}. As aging is gradually introduced, we
observe a descending slope emerging first at the leftmost part of the curve
(Fig. 5). It becomes increasingly manifest until the scaling law is
completely observed. The rate of the slope remains around $0.75$ in the
whole process. In this specific model, this value is independent of the
intensity of the aging effect. This scaling law clearly indicates that a
hierarchical structure is produced by the aging effect.

\subsection{Degree Correlation}

\label{Sec. 3. 3}

In the following we discuss the aging effect on the degree-degree correlation%
\cite{ass-mix} (or the mixing pattern, as it is sometimes called). For
convenience we shall call a node with $k$ degrees a $D-k$ node.

The degree correlation of nearest neighboring nodes is an important generic
property of networks. It can be quantified by the probability matrix $%
P\left( k,k_{nn}\right) $, i.e., the probability that a $D$-$k$ node is
connected with a $D-k_{nn}$ node. However, in reality, with the available
empirical data, a direct plot of $P\left( k,k_{nn}\right) $ often results in
a noisy picture difficult to interpret. An equivalent choice\cite{Internet1}
is to measure instead the nearest neighbors' average degree of the $D-k$
nodes, $\left\langle k_{nn}\right\rangle _k=\sum_{k_{nn}}k_{nn}P\left(
k,k_{nn}\right) $, as a function of $k$. Following Newman's idea\cite
{ass-mix}, if the high degree nodes in a network tend to connect to the low
(or other high) degree nodes, then we have a disassortative (or assortative)
mixing pattern; if there is no obvious bias, then we have a neutral mixing
pattern and $\left\langle k_{nn}\right\rangle _k=\left\langle
k^2\right\rangle /\left\langle k\right\rangle $, a value independent of $k$.

Before we discuss the correlation patterns with the aging effect, we provide
results of the BA model for comparison. In Ref. \cite{cal}, Krapivsky and
Redner have obtained in the BA model a useful characterization of
correlation, $N_{kl}\left( t\right) $, i.e., the number of $D-k$ nodes that
attach to a $D-l$ ancestor. Asymptotically, $N_{kl}\left( t\right)
\rightarrow tn_{kl}$, and 
\begin{equation}
n_{kl}=\frac{4\left( l-1\right) }{k\left( k+1\right) \left( k+l\right)
\left( k+l+1\right) \left( k+l+2\right) }+\frac{12\left( l-1\right) }{%
k\left( k+l-1\right) \left( k+l\right) \left( k+l+1\right) \left(
k+l+2\right) }.
\end{equation}
Here we consider undirected links and study $N_{kl}^{\prime }=N_{kl}+N_{lk}$%
, the number of $D-k$ nodes that are {\it linked with} a $D-l$ node.
Asymptotically 
\[
N_{kl}^{\prime }/t\rightarrow n_{kl}^{\prime }=n_{kl}+n_{lk}. 
\]
The probability that a nearest neighbor of a $D-k$ node is $D-k_{nn}$ is 
\[
\left( N_{k,k_{nn}}+N_{k_{nn},k}\right) /\sum_{k_{nn}}\left(
N_{k,k_{nn}}+N_{k_{nn},k}\right) 
\]
and the average degree of the nearest neighbors of the $D-k$ nodes is 
\begin{equation}
\left\langle k_{nn}\right\rangle _k=\frac{\sum_{k_{nn}}k_{nn}\left(
N_{k,k_{nn}}+N_{k_{nn},k}\right) }{\sum_{k_{nn}}\left(
N_{k,k_{nn}}+N_{k_{nn},k}\right) }\rightarrow \frac{\sum_{k_{nn}}k_{nn}%
\left( n_{k,k_{nn}}+n_{k_{nn},k}\right) }{\sum_{k_{nn}}\left(
n_{k,k_{nn}}+n_{k_{nn},k}\right) }.
\end{equation}
We show it approximately in Fig. 6(a) by taking the summation of $k_{nn}$ to 
$1.5\times 10^5$ and $2\times 10^6$ (with normalization satisfied)
respectively, in comparison with the simulation results at $N=100$, $1000$
and $10000$. It is found that nodes with large $k$ show no obvious biases in
their associations. But, there is a short disassortative mixing region when $%
k$ is relatively small.

Now we introduce a tunable aging effect. With the parameter $\beta $ taking
different values, the $\ln $-$\ln $ plots of $\left\langle
k_{nn}\right\rangle _k$ are shown in Fig. 7(a). When $k$ is relatively
small, they all descend linearly with approximately the same slope. This
indicates that in this region $\left\langle k_{nn}\right\rangle _k$ decays
as a power-law, $k^{-\lambda }$, and the exponent $\lambda $ is largely
independent of the intensity of the aging effect. But they show different
trends as $k$ increases. (1) For small values of $\beta $ ($\beta =0.01$ and 
$0.1$ in Fig. 7(a)), the curves become flatter as $k$ increases. (2) For
large $\beta $ ($\beta =0.4$ and $0.5$ in Fig. 7(a)), the power-law is
truncated with a fast decaying tail and (3) in the middle ground ($\beta
=0.2 $ and $0.3$ in Fig. 7(a)), the power-law is maintained. To conclude,
the aging effect, possibly above a certain intensity, leads to a
disassortative mixing pattern.

To explain the emergence of this mixing pattern, we directly study the
probability matrix $P\left( k,k_{nn}\right) $. In the BA model, Fig. 6(b)
shows two characteristic distributions of the probability that a nearest
neighbor of a $D-k$ node is $D-k_{nn}$. They are shown to be declining as
approximately $k_{nn}^{-2}$ for large $k_{nn}$, which is a sign of neutral
mixing since the probability of finding a $D-k_{nn}$ node in the network is
approximately $k_{nn}^{-3}$\cite{BA,cal}. It also leads to the conclusion
that $\left\langle k_{nn}\right\rangle $ will diverge for infinite network
size $N$ as $\ln N$, since the largest possible value of $k\sim N^{1/2}$\cite
{BA}, and approximately 
\begin{equation}
\left\langle k_{nn}\right\rangle \sim \frac{\int_1^{\sqrt{N}}k_{nn}\cdot
k_{nn}^{-2}dk_{nn}}{\int_1^{\sqrt{N}}k_{nn}^{-2}dk_{nn}}\sim \ln N.
\label{knnN}
\end{equation}
This trend is observed in Fig. 6(a). A three-dimensional drawing of the
matrix $P\left( k,k_{nn}\right) $ without aging can be found in Fig. 6(c).
When aging is considered, the effect is remarkable, as is shown in Fig. 7(b)
and 7(c). On the one hand, the probability distribution is changed from a
power law to a bell-shaped type (Fig. 7(b)); on the other hand, with $k$
growing, this bell-shaped distribution is shifting translationally, along
the negative direction of the axis of $k_{nn}$.

\subsection{Average distance between nodes}

\label{Sec. 3. 4}

Finally we study another fundamental topological feature of complex
networks, the average node-node distance, $\left\langle d\right\rangle $.
Here the distance between two selected nodes is defined as the number of
edges along the shortest path connecting them\cite{review1,review2,review3}.
In our simulation we calculate $\left\langle d\right\rangle $ as $%
\sum_{i=1}^N\left\langle d_i\right\rangle /N$, where $\left\langle
d_i\right\rangle $ is the average distance between the node $i$ and the rest
of the network.

As is mentioned in the Introduction, many complex networks show striking
small-world properties and have a relatively small value of $\left\langle
d\right\rangle $ compared with their large size. This effect is shared by
many models, including the Small-World model, the BA model and the random
network. However, with the aging effect strong enough, it is imaginable that
each node could only be connected with those introduced shortly before it.
Thus we may probably have a chain-like structure, with $\left\langle
d\right\rangle $ scaling linearly as $N$. In fact, as is shown below, this
depends on the choice of the decay factor.

(1) When $e^{-\beta \tau }$ is applied, we obtain a chain-like structure
even with vanishingly small $\beta $ ($\beta =0.01$ in Fig. 8(a)), and $%
\left\langle d\right\rangle $ increases linearly as $N$. However, a
chain-like structure does not necessarily mean a large value of $%
\left\langle d\right\rangle $. Here $\left\langle d\right\rangle $ is still
relatively small compared with $N$. (2) When $\tau ^{-\nu }$ is applied, we
observe a continuous transformation process that can be roughly identified
with three stages: The first stage is a small-world network with $%
\left\langle d\right\rangle \sim \ln N$ ($\nu =1$ in Fig. 8(b)); in the
second stage the network is similar to a hypercubic lattice with $%
\left\langle d\right\rangle \sim N^{1/D}$, where $D$ is the Euclidean
dimension ($\nu =2$ and $D\approx 2.73$ in Fig. 8(c)); finally, in the third
stage, the network evolves into a chain-like structure with $\left\langle
d\right\rangle \sim N$ (Fig. 8(d)). The difference between the two decay
factors becomes more manifest when we take into consideration the position
of each node and plot $\left\langle d_i\right\rangle $ as a function of $i$.
Fig. 9(a), obtained with the decay factor $e^{-0.01\tau }$, is just a sign
of a chain-like structure. By contrast, in Fig. 9(b) with the decay factor $%
\tau ^{-1}$, the nodes born earlier have shorter distance.

For each node, the probability that it is linked with a newly introduced
node is proportional to its degree and the decay factor, and the structure
of the generated network is decided by such a competition. A major
difference between the two decay factors, $e^{-\beta \tau }$ and $\tau
^{-\nu }$, is that $e^{-\beta \tau }$ decays much faster. With $e^{-\beta
\tau }$, the probability that two temporally distant nodes are connected is
so small that a chain-like structure will be produced even with vanishingly
small $\beta $. However, when $\tau ^{-\nu }$ is chosen, the result of the
competition depends on the parameter $\nu $, and that is why we have
observed a continuous transformation. Besides, it is worth mentioning that
actually the small-world property can be retained with $\nu $ in a
remarkably large region.

Finally, it is worth mentioning that, from the observation of an exponential
aging in the citation network and the present model, we may predict a linear
increase of the average distance between nodes (in the citation web) with
the number of nodes in the network. This is against the common expectation
of a small ''world behavior'', where a logarithmic increase is observed, but
it is certainly what follows from our work. We hope this conclusion will
stimulate further statistical measurements of citation networks.

\section{Summary}

\label{Sec. 4}

In network evolution, aging is a universal effect: After all, nothing is
perpetual. Here we find in citation networks that the aging effect may be
represented by an exponential decay factor, $e^{-\beta \tau }$, while this
particular measurement does not exclude the possibility that other evolving
networks may have different types of aging, for example, a power-law decay,
which is also studied and compared. It has been found that, as soon as the
preferential attachment is modified by such a factor, the produced network
will be significantly transformed, besides the change of the degree
distribution\cite{aging-pro1,aging-pro2}. In most cases we observe a
continuous transformation process by tuning the decay factor, while there
are also a few abrupt changes. (1) The network will be clustered even with
infinite network size, and the clustering coefficient varies greatly with
the intensity of the aging effect, i.e. it increases with $\beta $ linearly
for small values of $\beta $ and decays exponentially for large values of $%
\beta $. At the same time, the aging effect may also result in (2) a
hierarchical structure and (3) a disassortative degree-degree correlation,
and we observe how the corresponding scaling laws gradually emerge. (4)
Generally the aging effect will increase the average node-node distance in a
network, but the result depends on the chosen decay factor and the
intensity: If exponential decay is applied, the network appears like a
one-dimensional chain; but with power law decay a transformation process is
demonstrated, i.e., from a small-world, to a hypercubic lattice and to a
one-dimensional chain finally.

Presently there are plenty of problems worthy of further investigation. For
example, the influence of different choices of the decay factor, hidden
behind similar statistical properties, on network structure. In the present
research, the interesting disparities revealed, concerning the clustering
and the node distance, and probably with other aspects not yet identified,
are believed to bear significant meaning for empirical data acquisition. We
hope that the measurement conducted in the present work, about the citation
web, will be applied to more systems in future empirical studies. The effect
of aging on network structure observed in the present article also justifies
a parallel study of the effect on network function\cite{review1}, which
includes topics such as efficiency, information and disease transportation,
error and attack tolerance, and percolation features, etc.

\acknowledgements 

The authors would like to thank Konstantin Klemm for his help in providing
the original data used in Fig. 1. This work was supported by the National
Natural Science Foundation of China under Grant No. 10075025.

\appendix

\section{Calculation of the asymptotic clustering coefficient}

\label{Appendix A}

Because of the decay factor $e^{-\beta \tau }$, each node can only be linked
with a finite region, however large the whole network may actually be. We
can use this property to simplify the calculation of the asymptotic
clustering coefficient of a infinitely large network.

As is described in Sec. \ref{Sec. 2}, the model network is built as the
following: at each time step a single node is introduced, and then it is
attached to $m$ existing nodes. After the network is built, we calculate the
clustering coefficient of a randomly selected node, which is numbered as
node $0$. We number the node introduced $i$ time steps before it as node $-i$
and the node introduced $i$ time steps after it as node $+i$. As a result of
the aging effect, we can assume that each node of this network has the same
connectivity $2m$. Thus, the probability that two nodes $i$ and $j$ are
connected can be written as (a node cannot be linked with itself) 
\begin{eqnarray}
P_{link}\left( i,j\right) &=&m\frac{2m\left( e^{-\beta \left| i-j\right|
}-\delta _{ij}\right) }{\sum_{l=1}^\infty 2me^{-\beta l}}=m\frac{e^{-\beta
\left| i-j\right| }-\delta _{ij}}{\sum_{l=1}^\infty e^{-\beta l}}  \nonumber
\\
&=&m\left( e^\beta -1\right) \left( e^{-\beta \left| i-j\right| }-\delta
_{ij}\right) .  \label{link-prob}
\end{eqnarray}
The clustering coefficient of the node $0$, 
\begin{eqnarray*}
C\left( \beta \right) &=&\frac{\sum_{i=-\infty }^{+\infty }\sum_{j=-\infty
}^{+\infty }P_{link}\left( 0,i\right) P_{link}\left( 0,j\right)
P_{link}\left( i,j\right) }{2m\left( 2m-1\right) } \\
&=&\frac{m^3\left( e^\beta -1\right) ^3}{2m\left( 2m-1\right) }\left[
2\sum_{i=1}^\infty \sum_{j=1}^\infty e^{-\beta i}e^{-\beta j}\left(
e^{-\beta \left| i-j\right| }-\delta _{ij}\right) +2\sum_{i=1}^\infty
\sum_{j=1}^\infty e^{-\beta i}e^{-\beta j}e^{-\beta \left( i+j\right)
}\right] \\
&=&\frac{6m^3}{2m\left( 2m-1\right) }e^{-\beta }\frac{\left( 1-e^{-\beta
}\right) ^3}{\left( 1-e^{-2\beta }\right) ^2}.
\end{eqnarray*}
In fact this value does not depend on which node we select, and thus it can
be taken as the clustering coefficient of the network. As is shown in Fig.
3, this approximate calculation has a better fit with the simulation results
for large values of $\beta $. Actually, the difference mainly comes from Eq.
(\ref{link-prob}), in which we do not consider the degree difference of the
nodes.

\null\vskip0.2cm

\centerline{\bf Caption of figures} \vskip1cm

Fig. 1 (a) The scientific citation web formed by papers (nodes) and
citations (directed bonds). The open squares correspond to papers published
in each year between 1987 and 1998 and the solid squares correspond to
citations made in 1998 and referring to papers published in a given year\cite
{citationweb1}. The data have been extracted from the ISI\ database\cite
{citationweb2}. The average number of citations a paper published in a given
year received in 1998, which is actually a ratio of the other two curves, is
shown in diamonds. (b){\bf \ }The natural logarithm of the calculated
function $f\left( t\right) =\left[ 1/k\left( t\right) \right] dk\left(
t\right) /dt$ versus time step $t$ (solid circles). The linear fit (solid
line) corresponds to a exponential decay, $e^{-\beta t}$, with the exponent $%
\beta =0.28353$.

Fig. 2 The clustering coefficient as a function of system size $N$, in the
model described in text, with $\beta =0$ (squares), $10^{-3}$ (circles), $%
10^{-2}$ (upward triangles), $0.1$ (leftward triangles), $0.5$ (downward
triangles), and $7.7$ (diamonds), respectively. Each newly added node is
linked to $3$ existing nodes (also in all the following simulations). The
data points correspond to system sizes varying from $100$ to $20,000$, and
each is obtained as an average of many independent runs.

Fig. 3 The analytical result of the asymptotic clustering coefficient of an
infinitely large network, $C\left( \beta \right) $, as a function of $\beta $
(line), in comparison with the simulation (squares) results. Here $m=3$ and
the network size $N=20000$ in the simulation.

Fig. 4 In a $10,000$-node network with the decay factor chosen to be $\tau
^{-\nu }$, the ratio of the average clustering coefficient $C_2$ of the
second half and $C_1$ of the first half, as a function of $\nu $.

Fig. 5{\bf \ }The $\ln $-$\ln $ plot of the clustering coefficient $C\left(
k\right) $ versus the connectivity $k$, with $\beta =0.01$ (squares), $0.1$
(circles), $0.2$ (upward triangles), $0.3$ (downward triangles), $0.4$
(diamonds), $0.5$ (crosses), and the system size $N=10,000$. The solid line
is a power-law decay, $k^{-\gamma }$, with the exponent $\gamma =0.75$.

Fig. 6{\bf \ }Degree-degree correlations in the BA model without aging.
(a)Average degree $\left\langle k_{nn}\right\rangle $ of the neighboring
nodes of the $D-k$ nodes as a function of $k$. There is no aging effect and $%
\beta =0$. Squares, circles and upward triangles correspond to the
simulation results with system size $N=100$, $1000$ and $10000$
respectively. The dashed line and the solid line represent the theoretical
results with $k_{nn}$ up to $1.5\times 10^5$ and $2\times 10^6$
respectively. (b) Degree distributions of the nearest neighbors of $D-3$
nodes (squares) and $D-20$ nodes (circles) respectively. The dashed lines
are the corresponding theoretical results. The solid line with slope $-2$
serves as a guide to the eye. (c) The probability matrix $P\left(
k,k_{nn}\right) $. In both (b) and (c) system size $N=10,000$.

Fig. 7 Degree-degree correlations with aging effect. (a){\bf \ }The $\ln $-$%
\ln $ plot of $\left\langle k_{nn}\right\rangle _k$ versus $k$, at $\beta
=0.01$ (squares), $0.1$ (circles), $0.2$ (upward triangles), $0.3$ (downward
triangles), $0.4$ (diamonds), $0.5$ (leftward triangles), with system size $%
N=10,000$. The solid line corresponds to a power-law decay $k^{-\lambda }$,
with the exponent $\lambda =0.183$. (b) Degree distributions of the nearest
neighbors of $D-6$ nodes (circles) and $D-14$ nodes (squares) respectively.
(c) The probability matrix $P\left( k,k_{nn}\right) $. Both (b) and (c) are
obtained with $N=3,000$.

Fig. 8 Average distance between nodes as a function of network size $N$ with
different decay factors: (a) $f\left( \tau \right) =e^{-0.01\tau }$, and $%
\left\langle d\right\rangle \sim N$; (b) $f\left( \tau \right) =\tau ^{-1.5}$%
, and $\left\langle d\right\rangle \sim \ln N$; (c) $f\left( \tau \right)
=\tau ^{-2}$, and $\left\langle d\right\rangle \sim N^{-1/2.73}$; (c) $%
f\left( \tau \right) =\tau ^{-3}$, and $\left\langle d\right\rangle \sim N$.
The results are obtained with $m=2$.

Fig. 9 The average distance between each given node in the network and all
the other nodes, with the decay factor as (a) $f\left( \tau \right)
=e^{-0.01\tau }$ and (b) $f\left( \tau \right) =\tau ^{-1}$. Here $N=20000$
and $m=2$.

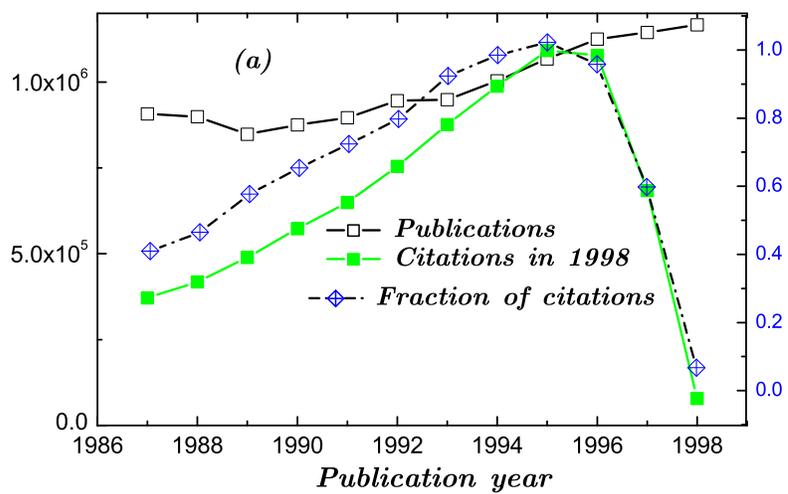
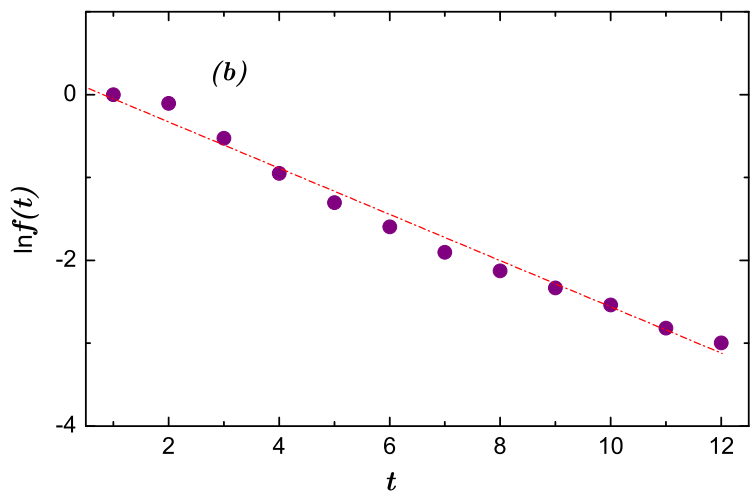
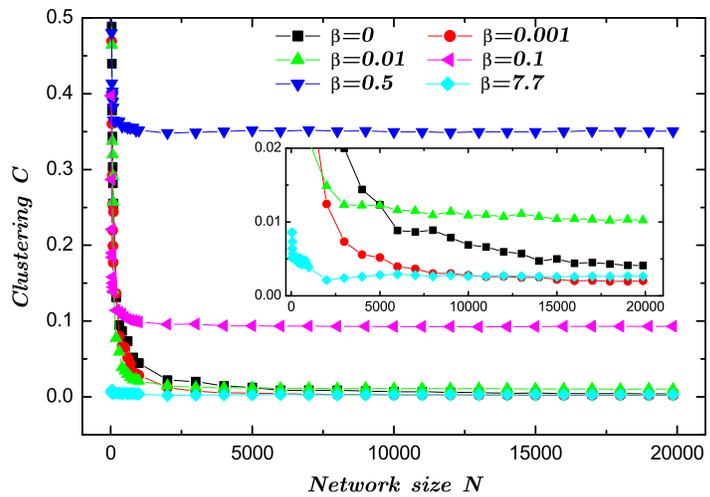
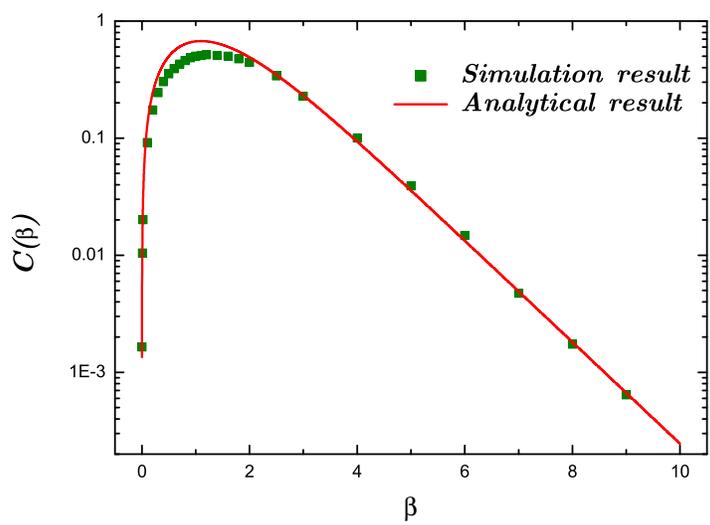
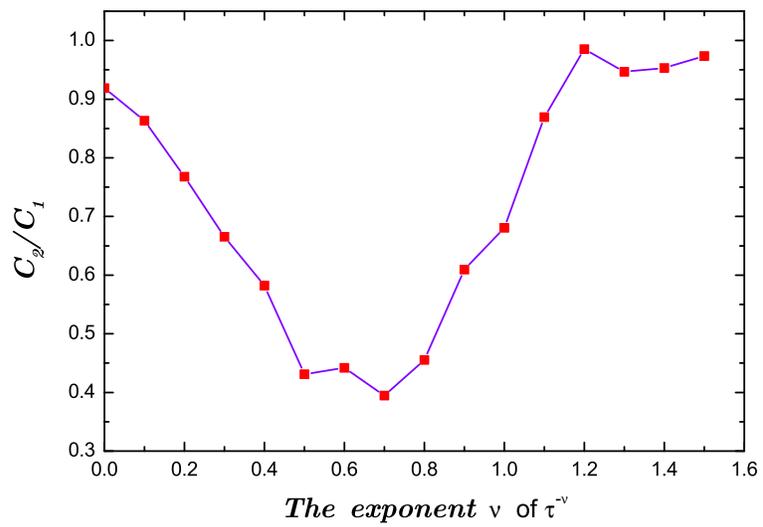
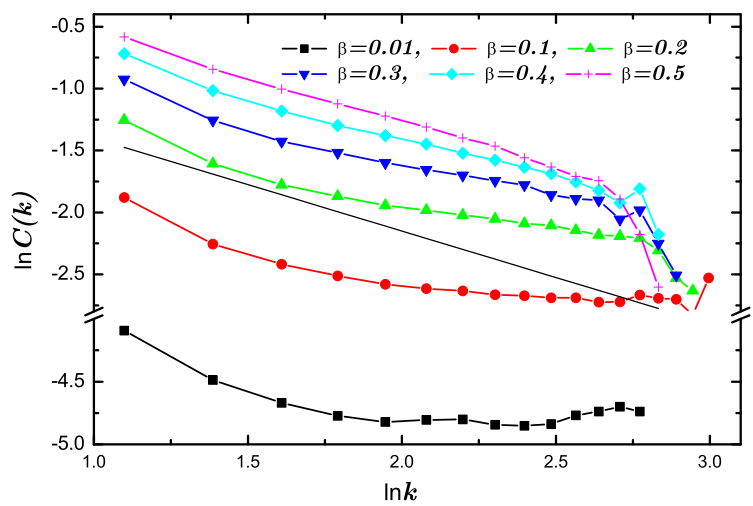
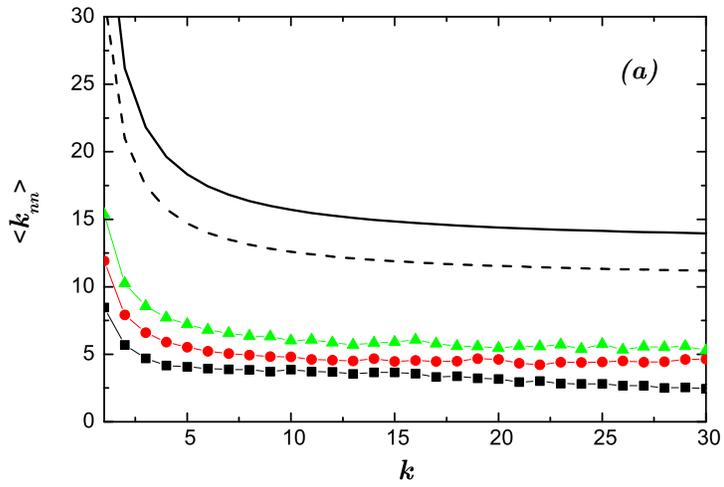
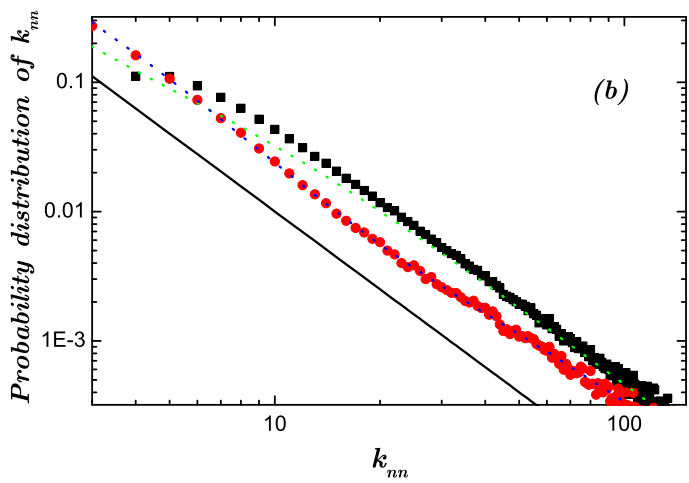
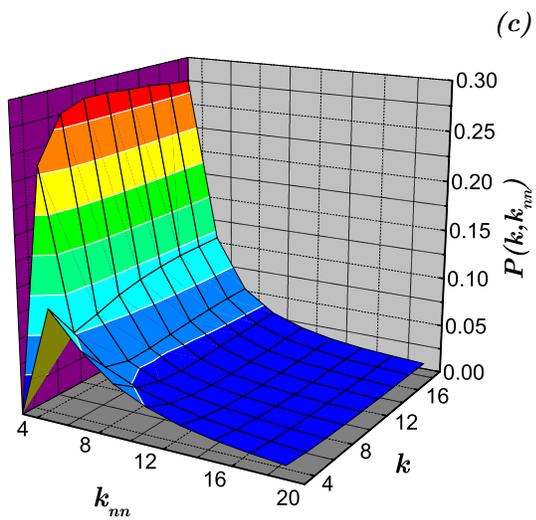
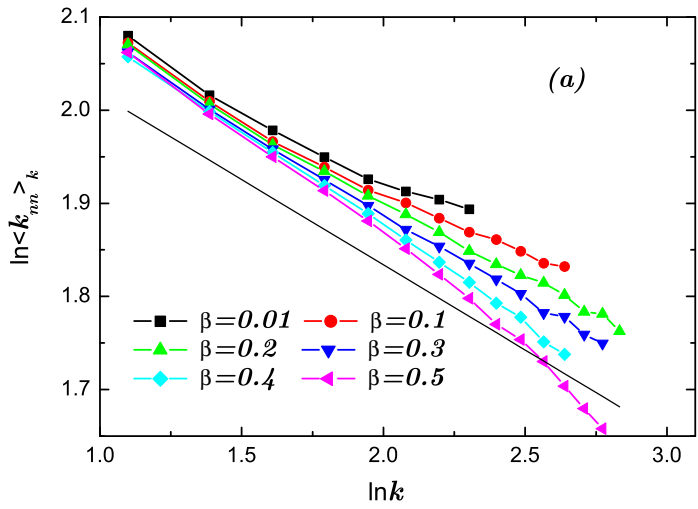
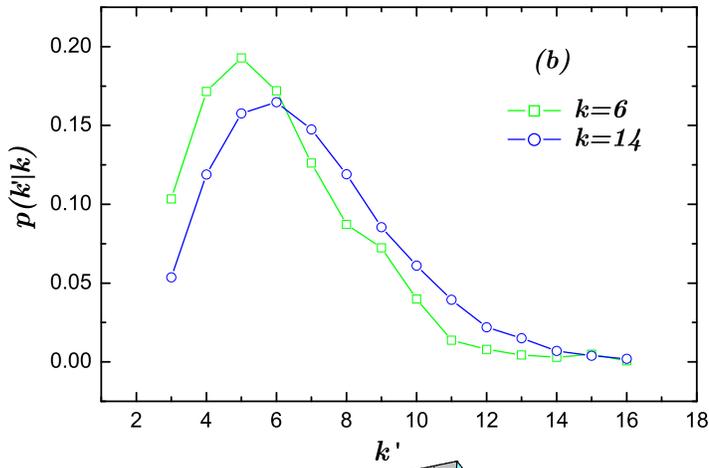
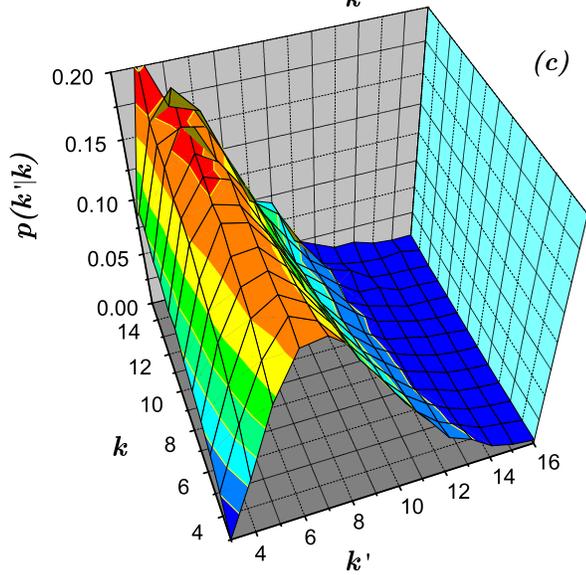
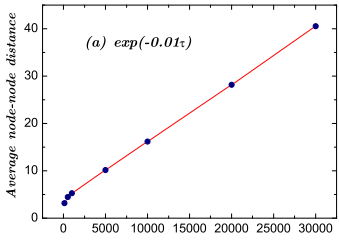
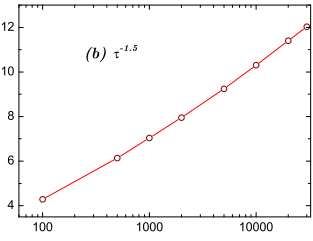
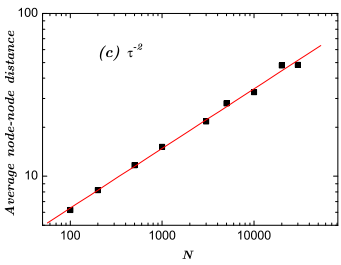
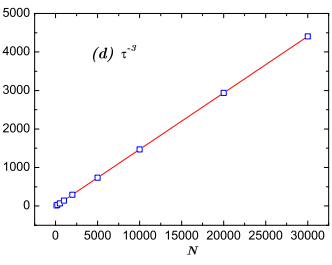
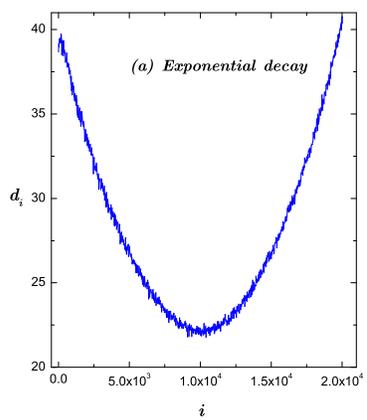
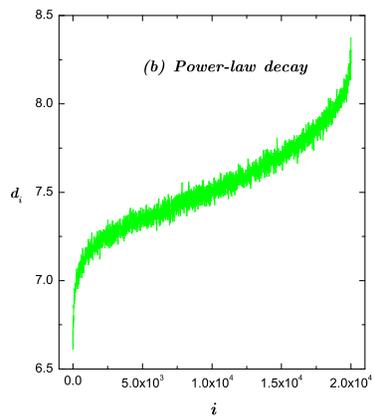
\end{document}